# Processes accompanying stimulated recombination of atoms

M.A. Kutlan

Institute for Particle & Nuclear Physics, Budapest, Hungary

kutlanma@gmail.com

The phenomenon of polarization of nuclei in the process of stimulated recombination of atoms in the field of circularly polarized laser radiation is considered. This effect is considered for the case of the proton-electron beams used in the method of electron cooling. An estimate is obtained for the maximum degree of polarization of the protons on components of the hyperfine structure of the 2s state of the hydrogen atom.

## 1. INTRODUCTION AND FORMULATION OF THE PROBLEM

Laser spectroscopy of hyperfine-structure (HFS) states is being applied in investigations of the properties of atoms and nuclei [1,2 and references therein]. The polarization of nuclei that arises in multiphoton resonance ionization of atoms was considered.

In connection with experiments on electron cooling of ions in storage rings [3] it is of interest to consider the inverse process - namely, stimulated recombination of atoms (ions) with subsequent population of hyperfine-structure (HFS) components. Effects involving stimulated recombination of atoms in proton - electron beams were considered in [4 - 6]. It was shown that under certain conditions, ensured by the method of electron cooling, the rate of stimulated recombination considerably exceeds the rate of spontaneous recombination. In view of this, it is natural to expect that in the case of stimulated recombination in the field of a circularly polarized wave it will be possible to observe effects involving the optical polarization of nuclei.

The process of the polarization of the nuclei occurs in two stages. In the initial state there are unpolarized proton and electron beams, propagating in the cooling part with equal average velocities. We shall consider the situation that arises at the end of the cooling, when, in the co-moving reference frame, the temperature of the protons in all their degrees of freedom is comparable to the longitudinal temperature of the electrons [7].

Parallel to these beams (along the z axis) is transmitted a laser wave, circularly polarized in the transverse (xy) plane and tuned to resonance with certain free-to-bound transitions. As a result of stimulated recombination in the field of this wave, selective population of certain states of an intermediate level of the bound system occurs. In particular, the role of this level could be played by the fine-structure components $3p_{3/2}$ and $3p_{1/2}$ of the hydrogen atom.

In the case when one wave is used, as a result of spontaneous transitions from the states $3p_{3/2,1/2}$ the HFS components of the 2s and 1s states of the atom become populated. The rate of these transitions is determined by the natural width $\gamma$ of the 3p states, but the degree of polarization of th nuclei in the final states is found to be negligibly small. It proportional to the factor $\Delta\varepsilon_{HFS}/\Delta E_{nm}$, where $\Delta\varepsilon_{HFS}$ is the energy of the hyperfine splitting of the appropriate states and $\Delta E_{nm}$ is the energy of the transition from the intermediate to the final state. As will be shown below, in the case of stimulated transitions, from the $3p_{3/2,1/2}$ sates to components of the HFS of the metastable level $2s_{1/2}$ in the field of a second resonance wave, the degree of polarization turns out to be much greater and reaches values ~ 1.

The degree of polarization of the nuclei can be calculated for three variants, differing in the type of polarization of the second wave and in the geometry of the experiment: a) a laser wave with linear (along the z axis) polarization and with propagation direction transverse to the beams; b) a wave with circular polarization in the xy plane, propagating along the beams (along the z axis); c) a wave with linear polarization, transmitted along the beams.

In principle, case (c) does not differ in any way from case (b), since linear polarization in the plane perpendicular to thez axis can be described by a superposition of two waves with circular polarizations in which the directions of rotation of the vector E in the waves are opposite. The difference then reduces entirely to the appropriate recalculation of the intensities of the waves.

We note that variant (a) is the least effective from the point of view of the rate of recombination of the atoms into the final states, since for this variant the volume of the region of interaction of the beams with the waves turns out to be substantially smaller. In addition, in view of the relatively high transverse temperature of the electrons in the beam, the Doppler shift $\Gamma_{D\perp}$ of the radiation in the directions perpendicular to the z axis is considerably greater than the natural width $\gamma$ of the $3p$ level. For these reasons, we shall confine ourselves henceforth to analyzing the case of two waves propagating along the beams [variant (b)].

The numerical values of the parameters of the hydrogen-atom states are taken from Ref.[7].

## 2. THE BASIC EQUATIONS

We define the degree of polarization of the nuclei of the recombined atoms by the expression

$$P = \frac{\Delta w}{w(M=1/2) - w(M=-1/2)}, \tag{1}$$

where the denominator is the total probability of formation of hydrogen atoms in all the HFS components of the $2s_{1/2}$ state.

If in all the expressions (4), (5), and (6) of [2,8] for the transition probabilities we retain only the terms corresponding to transitions via the intermediate level $3p_{3/2}$, it follows from the definition (1) that

$$P \approx 0.34 \frac{\Delta \varepsilon_{HFS} (dI/d\omega_2)_{\omega_{20}}}{I_{\omega_{20}}}. \tag{2}$$

We shall describe the frequency spectrum of the second wave by the model dependence

$$I_{\omega_2} = I_0 \frac{\Delta \omega_2}{2\pi} \frac{1}{(\omega_2 - \omega_0)^2 + \Delta \omega_2^2 / 4}, \tag{3 9}$$

where $\Delta \omega_2$ is the half-width of the distribution function and $\omega_0$ is the frequency at the maximum. For an upper bound on the degree of polarization of the nuclei we obtain

$$P_{max} \approx \frac{\Delta \varepsilon_{HFS}}{\Delta \omega_2}, \tag{4}$$

and, thus, in the framework of the approximations used, this quantity is not small.

The effectiveness of the method proposed in this paper for optical polarization of nuclei in beams is determined by the absolute rate of recombination of the atoms. For a correct estimate of this rate it is necessary to take into account the energy spread of the colliding particles, and also the pulsed character of the operation of the laser and storage ring. An expression for the matrix element for stimulated recombination from an s state of the continuous spectrum of the incident electron to a 3p state of the hydrogen atom was obtained in [6]. We shall confine ourselves here to the limiting case of this expression when $\varepsilon_{01}$, $\varepsilon_{02} \ll Ry$ ($Ry = m_e e^4 / 2\hbar^2 = 13.6 eV$):

$$\left| V_{p/n}^{(1)} \right|^2 \approx (2\pi)^2 \left( \frac{32 \times 9}{e^6} \right)^2 (eE_{01} a_0)^2 \frac{|C|^2}{4} \left( \frac{Ry}{\varepsilon} \right)^{1/2}, \tag{5}$$

where C is the normalization coefficient of the $\psi$ function of the electron in the continuous spectrum: $a_0 = \hbar^2 / m_e e^2$ is the first Bohr radius of the hydrogen atom; in the denominator of the numerical factor, e = 2.7 18 ... .

At the end of the cooling process the spectrum of the incident electrons in the center-of-mass frame of the colliding particles is given by the "flat" Maxwell distribution

$$F(\varepsilon_\|, \varepsilon_\perp) = \frac{1}{(\pi k T_\|)^{1/2}} \frac{1}{2kT_\perp} \exp\left(-\frac{\varepsilon_\|}{kT_\|}\right) \exp\left(-\frac{\varepsilon_\perp}{kT_\perp}\right) \varepsilon_\|^{-1/2}, \tag{6}$$

where the spread of the transverse velocities of the electrons is determined by the transverse temperature $T_\perp \sim T_c$ ($T_c$ is the temperature of the electron-beam cathode); the spread of the longitudinal velocities as a result of the potential acceleration of the electrons is substantially smaller and is characterized by a temperature $T_\| \approx T_c/(kT_c/4E_e)$, where $E_e$ is the kinetic energy of the electrons in the laboratory frame.

The rate of recombination of atoms, averaged over the distribution (6), is determined by the integral

$$\left\langle \frac{dN}{dt} \right\rangle \propto \iint \frac{\left|V^{(1)}_{p/n_1}\right|^2}{(\varepsilon - \varepsilon_{01})^2 + \tilde{\Gamma}^2_{n_1}/4} I_{\omega_{20}}(\varepsilon) F(\varepsilon_\|, \varepsilon_\perp) d\varepsilon_\perp d\varepsilon_\|, \tag{7}$$

in which the integrand can be represented by a product of three delta-like functions of the incident-electron energy $\varepsilon = \varepsilon_\| + \varepsilon_\perp$. The value of the integral depends on which of these functions is the sharpest. In a real situation, for typical parameters of the electron-cooling method, the conditions $kT_\perp \gg \Delta\omega_2 > kT_\| > \Gamma_n$) are realized. For the averaged rate of recombination of atoms into $2s_{1/2}$ states with a particular projection M of the nuclear spin we obtain (in the usual units)

$$\left\langle \frac{dN}{dt} \right\rangle = \frac{(2\pi)^3}{\hbar} \frac{1}{3} \left(\frac{3.1 \times 24}{e^6}\right)^2 \frac{(eE_{01}a_0)^3}{kT_\perp \tilde{\Gamma}_n \hbar \Delta\omega_2} n_e n_p V \left(\frac{l}{L}\right) \left(\frac{Ry}{\pi kT_\|}\right)^{1/2}$$
$$\times \left[ 5\left(\frac{\pi kT_\|}{\varepsilon_{02}}\right)^{1/2} \exp\left(-\frac{\varepsilon_{01}}{kT_\|}\right) \Phi\left(\left[\frac{\varepsilon_{01}}{kT_\|}\right]^{1/2}\right) + 4\left(\frac{\pi kT_\|}{\varepsilon_{02}}\right)^{1/2} \exp\left(-\frac{\varepsilon_{01}}{kT_\|}\right) \Phi\left(\left[\frac{\varepsilon_{02}}{kT_\|}\right]^{1/2}\right) \right] \frac{1}{Q}. \tag{8}$$

In (8) we have used the following notation: $n_e$ and $n_p$ are the concentrations of electrons and protons in the beams; V is the volume of the region of intersection of the waves with the beams; l is the length of that part of the storage ring in which cooling of the protons occurs; L is the perimeter of the storage ring; $\Phi(x)$ is the probability integral; $Q = T/\tau$ is the off-duty factor of the pulsed laser creating the first wave; $E_{02}$ is the amplitude of the field intensity in the second wave.

We draw attention to a number of features of the expression (8):

1 ) The quantity $\left\langle \frac{dN}{dt} \right\rangle$ is proportional to the volume V of the region of interaction of the beams with the waves, and for this reason a scheme with parallel transmission of both waves along the beams is preferable.

2) The optimal conditions for observation of the effects under consideration arise when the ionization width and field width of the intermediate level 3p are of the order of the width $\Gamma_n$ : $\Gamma_i$ $\Gamma_f \sim \Gamma_n = \gamma + \Gamma_{D\parallel}$. This imposes upper limits on the admissible values of the electric-field intensities in the waves (for estimates, see below).

3) The condition for resonance of the stimulated recombination requires that the amount by which the energy $\hbar\omega_1$ of a quantum exceeds the threshold for ionization from the intermediate level be very small: $\varepsilon_{01}, \varepsilon_{02} \ll kT$. Otherwise, the effect is found to be exponentially small. But since, in a real situation, $\Delta\varepsilon_{HFS} \gg kT_\parallel$, only one term operates effectively in the square brackets in Eq. (8). In other words, the recombination of atoms proceeds in practice via one particular component of the fine structure of the 3p level (under the condition $\Delta\omega_1 \ll \Delta\varepsilon_{FS}$.

We shall formulate the conditions and obtain estimates for the admissible values of the field intensities in the waves. An expression for the photoionization width of the 3p level was found in [6]. In ionization to the edge of the absorption band ($\varepsilon_0 / kT_\parallel \to 0$) the formula for $\Gamma_i$ takes the form

$$\Gamma_i = \frac{\pi(eE_{01})^2 ca_0^5 m_e c^2 (2m_e c^2 Ry)^{1/2}}{(\hbar c)^4}.$$

The condition $\Gamma_n \geq \Gamma_i$ can be formulated in the form of the inequality

$$eE_{01} \leq (\hbar c)^2 \left[ \frac{\Gamma_n}{\pi ca_0^5 m_e c^2 (2m_e c^2 Ry)^{1/2}} \right]^{1/2}. \tag{9}$$

Hence, $E_{01} \leq 2 \times 10^5 V/cm$ (the tabulated value of $\gamma$ is $\gamma = 30 MHz$, and the Doppler width corresponding to the longitudinal temperature $kT_\parallel \sim 10^{-6} eV$ is estimated as $\Gamma_{D\parallel} \approx 1.4 \times 10^{-7} eV$. Such field intensities can be achieved by means of pulsed lasers with tunable frequency.

The field width of the intermediate level 3p in the field of the second wave is given, to within a numerical factor of order unity, by the formula

$$\Gamma_i = \frac{(2\pi)^2}{\hbar} \alpha a_0^2 I_{\omega_{20}} \sim \frac{(2\pi)^2}{\hbar} \alpha a_0^2 \frac{I}{\Delta\omega_2}. \tag{10}$$

Here the condition $\Gamma_n \geq \Gamma_f$ acquires the form

$$eE_{02} \leq \left[\frac{\hbar\Gamma_n \hbar\Delta\omega_2}{\pi a_0^2}\right]^{1/2}. \quad (11)$$

If for the characteristic spectral width Am, in the second wave we take $\Delta\omega_2 \sim \Delta\varepsilon_{HFS}$, then from (11) we obtain the bound . $E_{02} \leq 30 V/cm$.

## 3. CONCLUSION, ESTIMATES

In this section we obtain numerical estimates for the effects considered in the paper, using data on electron cooling of protons from [4]. When the electron energy is $E_e \leq 30 KeV$, the energy of protons with the same average velocity as the electrons is $E_p = E_e(M/m_e) = 100 MeV$. In the co-moving reference frame, the spread of the transverse velocities of the electrons is determined by the temperature $kT_\perp \approx kT_c \approx 0.2 eV$, and the spread of the longitudinal velocities is determined by the temperature $kT_\| \sim 10^{-6} eV$.

The length $l$ of the region in which the cooling of the protons occurs is $l \approx 2m$, and the length of the storage ring is $L \approx 200m$; the volume of the region of interaction of the beams with the waves is $V = (\pi d^2/4)l \approx 160 cm^3$ (d = 1 cm is the diameter of the cross section of the first wave); $n_e = n_p \approx 10^8 cm^{-3}$.

For the estimate, -in (8) we set $\hbar\Delta\omega_2 \approx \Delta\varepsilon_{HFS} \approx 1.2\times10^{-7} eV$, $\tilde{\Gamma}_n \approx \Gamma_n \approx 1.4\times10^{-7} eV$, $E_{02} = 30 V/cm$, and $E_{01} = 2\times10^5 V/cm$ (when the diameter of the light wave emerging from the laser is $d \approx 1 cm$ and the pulse duration is $\tau \approx 10^{-8} sec$, this gives a pulse energy $\upsilon = 10 KHz$. When the laser has pulse frequency $\nu$ = 10 kHz we obtain an off-duty factor $Q = 10^4$.

Retaining for the estimate just the first term in (8), in which we set $\varepsilon_{01} \approx 0$, we obtain $\left\langle\frac{dN}{dt}\right\rangle \approx 10^2 sec^{-1}$. This estimate has been found under the assumption that the proton beam continuously enters the region of electron cooling. If the size of a proton bunch is of the order of the length of the cooling region, this estimate can be increased by two orders of magnitude. The other aspects could be found in [9-62].

The proposed scheme of optical polarization gives us the possibility of populating the metastable $2s_{1/2}$ level of the hydrogen atom. Subsequent ionization of this state makes it possible to obtain a polarized proton beam. The estimates found in the paper for the degree of polarization of the nuclei are also valid, of course, in the case when the first wave is linearly polarized in the plane perpendicular to the *z* axis and the second wave is circularly polarized in the *xy* plane.